\documentclass[a4paper,twocolumn,showpacs,preprintnumbers,amsmath,amssymb]{revtex4}
\usepackage{graphicx}% Include figure files
\usepackage{dcolumn}% Align table columns on decimal point
\usepackage{bm}% bold math
\usepackage{geometry}
\usepackage{amsmath}

\begin{document}

%\preprint{APS/123-QED}

\title{Complexity of hierarchical ensembles}
\author{A.I.~Olemskoi}\email{alex@ufn.ru}, \author{S.V.~Kokhan}, \author{V.I.~Ostrik}
\affiliation{Institute of Applied Physics, Ukraine National Academy of Science,
40030 Sumy, Ukraine}

\date{\today}

\begin{abstract}
Within the framework of generalized combinatorial approach, complexity is
determined as a disorder measure for hierarchical statistical ensembles related
to Cayley trees possessing arbitrary branching and number of levels. With
strengthening hierarchical coupling, the complexity is shown to increase
monotonically to the limit value that grows with tree branching. In contrast to
the temperature dependence of thermodynamic entropy, the
 complexity is reduced by the variance of hierarchical statistical ensemble
if the branching exponent does not exceed the gold mean. Time dependencies are
found for both the probability distribution over ensemble states and the
related complexity. The latter is found explicitly for self-similar ensemble
and generalized for arbitrary hierarchical trees.
\end{abstract}

 \maketitle

\section{Introduction}\label{sec:1}

\begin{figure}
\centering
\includegraphics[width=80mm]{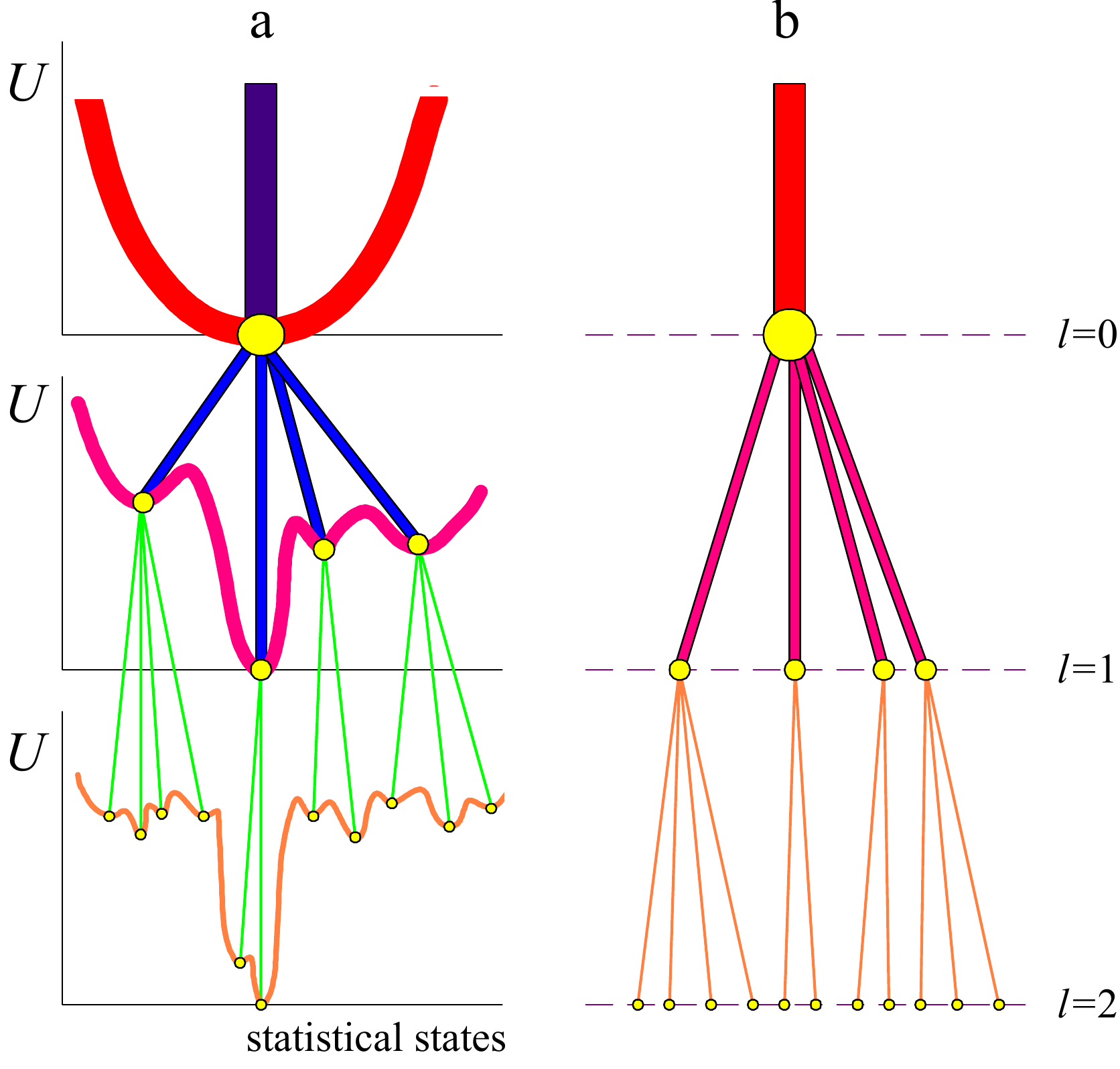}
\caption{Characteristic form of the internal energy landscape (a) and related
hierarchical tree (b) of complex system \cite{15}.}\label{fig0}
\end{figure}

Despite a daily appearance of hierarchy in society and comprehension of its
role in physical, biological, economical and other complex systems \cite
{1}-\cite {5} theory of hierarchically constrained statistical ensembles has
been developed only at description of dynamics of spin glasses \cite {6,7}.
Formal basis of this theory is that states of hierarchically constrained
objects are related to an ultrametric space whose geometrical image is the
Cayley tree with nodes corresponding to statistical (sub)ensembles \cite {8}.
The relaxation of hierarchical structures had been considered first \cite {10a}
as a diffusion process on either uniformly or randomly multifurcating trees
characterized by a diversity being a measure of the tree's complexity \cite
{9a}. Consequent study of the hierarchical ensembles has shown \cite{10} their
evolution was reduced to anomalous diffusion process in ultrametric space that
arrives at a steady-state distribution over hierarchical levels, which is a
Tsallis power law inherent in non-extensive systems \cite{11}. Principle
peculiarity of hierarchical systems consists in the splitting of each
statistical ensemble residing on a given level into a number of smaller
subensembles with passage onto the lower level; subsequent descent down the
tree results in more small subensembles (see Fig. \ref{fig0}).

From statistical point of view the set of above subensembles is characterized
by the complexity, whose value determines disorder of the hierarchical coupling
-- in analogy with the entropy of thermodynamic systems. Formally, above
quantities are defined equally, however{\bf,} their physical nature is
absolutely different: if the entropy characterizes disorder in distribution of
primitive structural units (for example, atoms), at the complexity definition
their role is played by subensembles into which the whole statistical ensemble
is subdivided. Consideration of self-similar hierarchically constrained
ensembles within the framework of generalized combinatorial approach shows
\cite {9} that with strengthening hierarchical coupling the complexity
increases monotonically to a certain boundary value; the latter decays with
both the growth of the variance of this coupling and the reduction of a
branching exponent of hierarchical tree. If the last of pointed dependencies is
obvious (indeed, non-branching tree does not possess a complexity), the
reduction of the boundary complexity with hierarchical ensemble scattering is
meant to be abnormal, as in the conventional thermodynamic ensembles the
entropy always increases with temperature. Obviously, with branching change,
promoting transformation of hierarchical ensemble into usual one, the pointed
anomaly should be weakened and the complexity decrease with the ensemble
scattering will pass into its increase.

The present work is based on generalization of the statistical approach \cite
{9} which allows, in particular, to establish the connection between the
entropy of simple systems and the complexity of hierarchical ensembles. To
avoid a misunderstanding, it is worth to note that the complexity introduced in
Ref.\cite {9a} characterizes a diversity of hierarchical tree's themselves,
whereas we are aimed to consider the disorder of hierarchically constrained
ensembles in their statistical distribution over such trees. Along this line,
in Section \ref{sec:2} initial statements are given to define the distribution
over states of self-similar statistical ensemble and the complexity of
arbitrary hierarchical system. Main results are obtained in Section \ref{sec:3}
for the complexity definition within continuum approach, where the main
contribution is shown to be given by the deepest levels of hierarchy.
Accounting for the discrete character of distribution in Section \ref{sec:4}
shows that continuum approach derives semi-quantitative results already for
several levels of hierarchy. Section \ref{sec:5}, where the complexity
definition is generalized for arbitrary hierarchical ensemble, is devoted to
discussion of the obtained results. Appendix contains details of generalized
combinatorial approach being a basis of our consideration.

\section{Main statements}\label{sec:2}

Generally, the behavior of a complex system is determined by the cluster
structure of the whole set of hierarchical levels, however, the property of
self-similarity enables one to consider a typical cluster and the level number
only. As a result, the system state is described by the probability
$p_l=p_l(t)$ to occupy the hierarchical level $l$ whose distribution obeys the
nonlinear Fokker-Planck equation \cite {10}
\begin{equation}
\tau_0\dot p_l = - {\partial\over\partial l} \left (\epsilon p_l^Q+D {\partial
p_l \over\partial l} \right). \label {1}
\end{equation}
Here, the level number $l\gg 1$ is supposed to be large to justify the use of
the continuum approach, the overpoint denotes the derivative with respect to
the time $t$ whose microscopic scale is $\tau_0$, and the diffusion coefficient
$D$ represents the second moment of the intensity of transitions between
microscopic states. Unlike the behavior of simple systems which is determined
by the linear drift corresponding to the first moment of this intensity, the
principle role in behavior of hierarchical ensembles plays the nonlinear term,
fixed by both the factor $\epsilon>0$ and the exponent $Q\in [1,2]$. At the
initial stage $t\ll \tau_d$, $\tau_d\equiv (\epsilon^{Q-2}/D^{Q-1})l^Q\tau_0$,
the contribution of diffusion is negligible and the characteristic scale of
hierarchy increases with time according to the power law
$l_c=Q^{1/Q}(t/\tau_0)^{1/Q}$, whereas the probability density
$p_l(t)=(l/Q\epsilon)^{1/(Q-1)}(t/\tau_0)^{-1/(Q-1)}$ decays hyperbolically
with time. Transition to the diffusion regime $\tau_d\sim t\ll\tau$,
$\tau\equiv l^2\tau_0$ leads to transformation of the time dependence $l_c(t)$
into the usual root form $l_c=\sqrt{2(t/\tau_0)}$, while the probability
density decays according to the same hyperbolic law. This law is appeared to be
always inherent in hierarchical ensembles, being not only self-similar but also
arbitrary ones \cite {9a}.

With growth of the time to macroscopic values $t\gg\tau $, the probability
distribution takes the stationary Tsallis form \cite {11}
\begin{equation}
\begin{split}
 &p_l =\left [p_0 ^ {-(Q-1)} + \frac {Q-1} {\Delta} l\right] ^ {-{1\over
Q-1}};\\ &p_0\equiv \left (\frac {2-Q} {\Delta} \right) ^ {\frac {1} {2-Q}},
\quad\Delta\equiv D/\epsilon. \label {2}
\end{split}
\end{equation}
The probability (\ref{2}) increases monotonically with $l$ decrease, i.e., with
growth of hierarchical cluster, reaching the maximum value $p_0$ on the top
level $l=0$ related to the whole system. Growth of the variance $\Delta\equiv
D/\epsilon$ expands considerably the stationary distribution over hierarchical
levels. Characteristically, in the limit $\Delta\gg 1$ the distribution
(\ref{2}) differs slightly from exponential one on high levels
$l\ll\Delta^{1/(2-Q)}$, however, with passing onto deeper levels the
distribution tail becomes power-law. Study of possible types of hierarchical
coupling \cite {12} has shown the distribution (\ref{2}) is inherent in
statistical systems related to a self-similar ultrametric space with fractal
dimension $d=(Q-1)/(2-Q)$.

\begin{figure}[!htb]
\centering
\includegraphics[width=80mm]{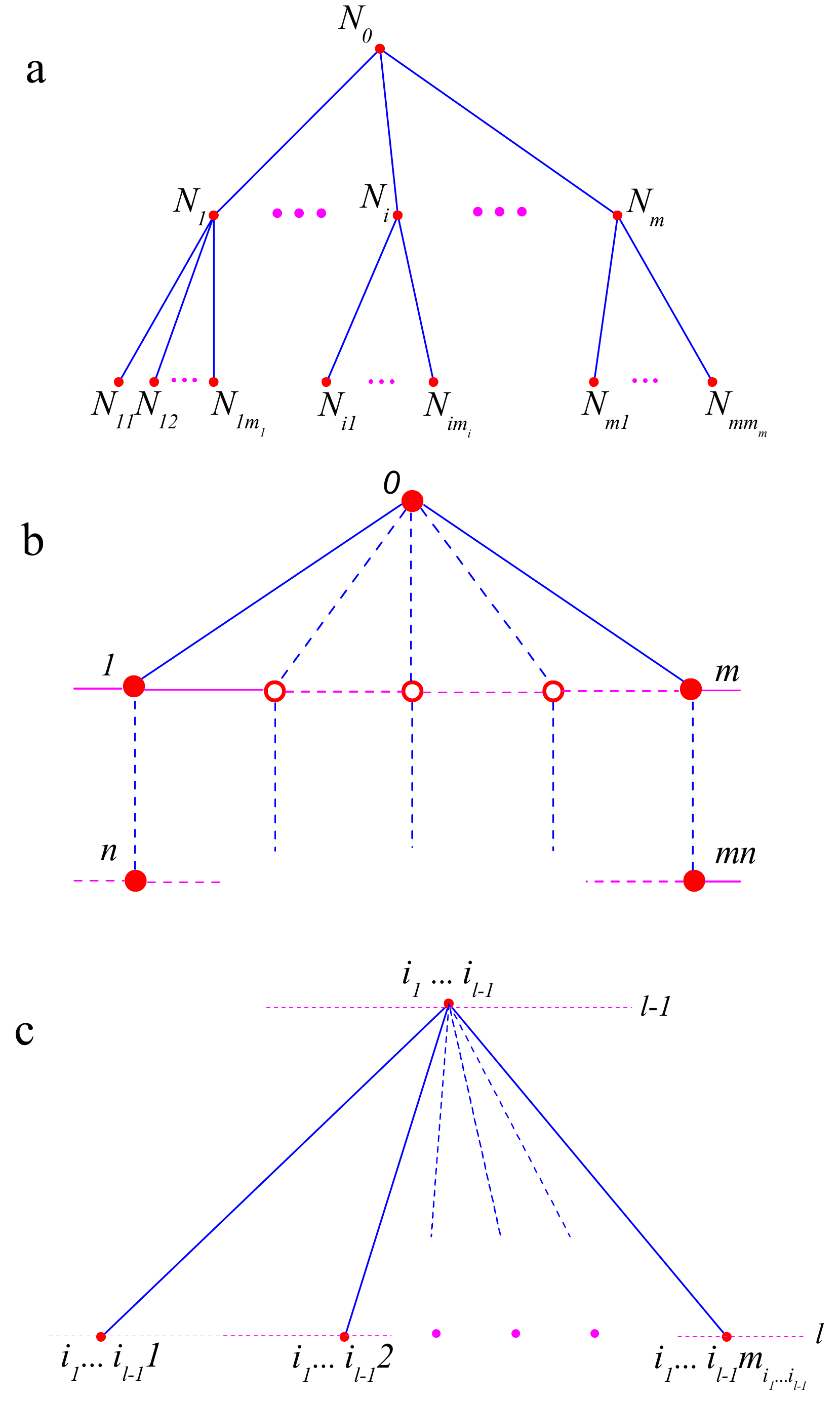}
\caption{ a) Numbers of states on nearest levels of typical hierarchical
cluster; b) trivial hierarchical tree; c) parameterization of nodes on nearest
levels.} \label{fig1}
\end{figure}

The statistical theory of self-similar hierarchical ensembles is based on
generalization of both logarithmic and exponential functions given by
expressions (\ref{A1}), being type of deformed exponential distribution
(\ref{2}). As shown in Appendix, the additivity condition of the logarithm
function is provided by the related deformation rules for the
multiplication/division operations, essentially changing the form of the usual
multinomial coefficients which are the basis of combinatorial formalism of
statistical physics \cite{13}. Making use of this formalism shows \cite{9} that
connection between complexities of the nearest hierarchical levels is expressed
by equality (\ref{A10}). In accordance with Fig.\ref{fig1}à
we suppose here that $N_0$ states of the top level are distributed over the $m$
groups on the lower level (labeled by $i$), each of which contains nodes $ij$
on the bottom level of a hierarchical cluster. If the group $i$ possesses $N_i$
occupied states and there are $N_{ij}$ occupied states within each set $ij$,
then the corresponding probabilities take the values $p_i=N_{i}/N$,
$p_{ij}=N_{ij}/N$, where $N$ is the total number of statistical states of
hierarchical ensemble. Note that $N$ is not reduced to value $N_0\ll N$ related
to the given cluster. Because of the obvious condition
$N_i=\sum_{j=1}^{m_i}N_{ij}$, the above probabilities are connected by the
equality
\begin {equation}
p_i =\sum_{j=1}^{m_i}p_{ij}.
 \label {4}
\end {equation}
As a result, expression (\ref{A10}), accompanied with Eqs. (\ref{A7})
(\ref{A8}), takes the form
\begin{equation}
\begin{split}
C_Q\left (p_{11}, \dots, p _ {mm_m} \right) &=C_Q\left (p _ {1}, \dots, p _ {m}
\right) +\\ &\sum\limits _ {i=1} ^ {m} p _ {i} ^Q C_Q\left (\frac {p _ {i1}}
{p_i}, \dots, \frac {p _ {im_i}} {p_i} \right).
 \label {3}
\end{split}
\end{equation}
Here, the last factor represents relative complexity defined by the Tsallis
expression
\begin {equation}
C_Q\left (\frac {p _ {i1}} {p_i}, \dots, \frac {p _ {im_i}} {p_i} \right) =
(Q-1) ^ {-1} \sum _ {j=1} ^ {m_i} \left [\frac {p _ {ij}} {p_i}-\left (\frac {p
_ {ij}} {p_i} \right) ^Q\right].
 \label {5}
\end {equation}
Utilizing this formula we express the connection between the complexities of
nearest hierarchical levels in terms of the related probability distributions
over statistical states:
\begin {equation}
\begin{split}
C_Q\left(p_{11},\dots,p_{mm_m}\right)&-C_Q\left(p_{1},\dots,p_{m}\right)=\\
&\frac{1}{Q-1}\sum_{i=1}^{m}\sum_{j=1}^{m_i}p_{ij}\left(p_i^{Q-1}-p_{ij}^{Q-1}
\right).
 \label {6}
 \end{split}
\end {equation}

Above consideration concerns a typical hierarchical cluster depicted in
Fig.\ref{fig1}à. To consider the whole statistical ensemble consisting of such
clusters, let us suppose that statistical states are distributed over
microcanonical (sub)ensembles with probabilities and corresponding complexities
defined by the level number $l$ only: $\{p_{ij}\}\Rightarrow p_{l+1}$,
$\{p_{i}\}\Rightarrow p_{l}$; $C_Q\left(p_{11},\dots,p_{mm_m}\right)
\Rightarrow C(l+1)$, $C_Q\left(p_{1},\dots,p_{m}\right)\Rightarrow C(l)$. As a
result, relation (\ref{6}) takes the simplest form
\begin {equation}
C(l+1)-C (l)=\frac{M_{l}}{Q-1}p_{l+1}\left(p_{l}^{Q-1}-p_{l+1}^{Q-1}\right).
 \label {7}
\end {equation}
For self-similar ensembles node numbers are distributed over levels $l=0,1,
\dots, n $ according to the power law \cite{12}
\begin {equation}
M_l = (l+1)^a
 \label {8}
\end {equation}
with an exponent $a>1$. In the limit $l\gg1$, the condition (\ref {4}) takes
the form
\begin {equation}
M_{l}p_{l}=M_{l+1}p_{l+1}.
 \label {9}
\end {equation}
Then, Eqs. (\ref{8}), (\ref{9}) accompanied with asymptotic behavior of
distribution (\ref{2}) show the self-similarity condition is fulfilled for
hierarchical trees related to the branching exponent
\begin {equation}
a=\frac{1}{Q-1}.
 \label {10}
\end {equation}

Stationary probability distributions over states of the nearest levels are
known to be connected by the recurrence relation \cite {10}
\begin {equation}
p_{l+1}-p_l =-p_l^Q/\Delta, \quad l=0,1, \dots, n.
 \label {11}
\end {equation}
In the limit $l\gg1$, this relation is reduced to the steady-state
Fokker-Planck equation (\ref{1}).

\section{Continuum approach}\label{sec:3}

Assuming that total number of levels is $n\to\infty $, let us find the
complexity of hierarchical ensemble where the main contribution is given by
deep levels with $l\gg 1$. Here, the statistical state probabilities of nearest
levels are connected by relations
\begin {equation}
\begin{split}
p_{l+1}^{Q-1}-p_l^{Q-1} &\simeq\frac{\rm d}{{\rm d}l}p_{l+1}^{Q-1} =\\
&(Q-1)p_{l+1}^{Q-2} \frac{{\rm d}p_{l+1}}{{\rm d}l} \simeq -\frac{Q-1}{\Delta}
p_{l+1}^{2(Q-1)}, \label {12}
\end{split}
\end{equation}
where the last equality is written with accounting for equation (\ref{11})
taken in the continuum limit. Then, substitution of Eq.(\ref{12}) into
Eq.(\ref{7}) gives
\begin {equation}
C (l+1)-C (l) \simeq\frac {M _ {l}} {\Delta} p _ {l+1} ^ {2Q-1}.
 \label {13}
\end{equation}
Rewriting $C (l+1)-C (l)$ in the continuum limit as ${\rm d} C / {\rm d}
  l $ and taking into account Eqs. (\ref{2}), (\ref{8}) allow one to reduce the
  difference equation (\ref{13}) to the following differential equation,
\begin {equation} \frac {{\rm d} C} {{\rm d} l} = \frac {l^a} {\Delta} \left
[p_0 ^ {-(Q-1)} + \frac {Q-1} {\Delta} ~l\right] ^ {-{2Q-1\over Q-1}}, \label
{14}
\end{equation}
where $l+1$ is replaced with $l$.

In the limiting case $a\to 1$, solution of Eq.(\ref {14}) leads to the
dependence
\begin{equation}
C (n) = \frac {\Delta} {Q} p_0\left [1-\left (\frac {p_n} {p_0} \right)
^Q\right]-np_n^Q.
 \label {15}
\end{equation}
Here, the complexity appears to be the monotonically increasing function of the
level number $n$. It has the initial value $C(0)=0$ and the maximum value
\begin {equation}
C (\infty) = \frac {(2-Q) ^ {\frac {1} {2-Q}}} {Q} \Delta ^ {-\frac {Q-1}
{2-Q}},
 \label {16}
\end{equation}
decaying with growth of both the non-extensivity parameter $Q $ and the
variance $\Delta$. Remarkably, the complexity obtained here behaves similarly
to that found in our recent work \cite {9}. However, the analogous expressions
(16)-(18) appearing in Ref. \cite{9} are much more complicated than the
equalities (\ref {14})-(\ref {16}) derived in this section. It is accounted for
that continuum approach has been already used in Ref.\cite{9} at estimation of
the relative entropy (\ref {5}), whereas in the present study we pass to
continuum limit at a later stage, while defining probability difference
(\ref{12}) to estimate the complexity difference (\ref {7}). At the same time,
no approximations have been imposed to express the entropy (\ref {5}). The
pointed distinction resulted in the exponent of the distribution function
derived in Ref.\cite{9} (see the r.h.s. of Eq.(16) therein) to exceed the
exponent value in Eq.(\ref {14}) of this section by the magnitude $Q-1<1$. This
fact, however, does not play a principle role within our approach.

At arbitrary values $a$ of the branching exponent, solution of equation
(\ref{14}) is expressed by means of the hypergeometric function:
\begin{equation}
\begin{split}
C(n) = \frac {p_0 ^ {2Q-1}} {\Delta} \frac {n ^ {1+a}} {1+a} ~&F\left (\frac
{2Q-1} {Q-1}, 1+a; 2+a;-\nu n\right),\\ &\nu\equiv\frac {Q-1} {\Delta} p_0 ^
{Q-1}.
 \label {17}
 \end{split}
\end{equation}
The analysis of r.h.s. of Eq.(\ref{14}) shows that its solution holds finite
values in the limit $n\to\infty $ if the exponent $a$ does not exceed the
maximum value
\begin{equation}
a _ {max} = \frac {Q} {Q-1}.
 \label {18}
\end{equation}
As shown above, the self-similarity condition leads to the lower value
(\ref{10}) for $a$, at which odd arguments of the hypergeometric function
coincide and one arrives at the binomial form \cite {14}
\begin{equation}
F\left (\frac {2Q-1} {Q-1}, \frac {Q} {Q-1}; \frac {2Q-1} {Q-1};-\nu n\right) =
\left (1 +\nu n\right) ^ {-\frac {Q} {Q-1}}.
 \label {19}
\end{equation}
As a result, the complexity (\ref {17}) is expressed by the simple equation
\footnote{We did not manage to find the elementary solution (\ref {20}) without
using the hypergeometric function.}
\begin{equation}
C (n) = \frac {Q-1} {Q} \frac {p_0 ^ {Q-1}} {\Delta} p_n^Qn ^ {\frac {Q}
{Q-1}}.
 \label {20}
\end {equation}
As for the degenerate tree ($a\to 1$), the complexity increases monotonically
with the number of hierarchical levels to the maximum value
\begin{equation}
C(\infty) = \frac {(2-Q) ^ {\frac {Q-1} {2-Q}}} {Q (Q-1) ^ {\frac {1} {Q-1}}}
\Delta ^ {\frac {Q + | Q_ - |} {(Q-1) (2-Q)} (Q _ +-Q)}, \label {21}
\end{equation}
where roots $Q _ {\pm} \equiv (1\pm\sqrt {5})/2$ represent the gold mean. With
growth of the non-extensivity parameter $Q$, the maximum complexity decreases
monotonically from infinite value at $Q=1$ to zero at $Q=2$. However, unlike
Eq.(\ref {16}), the complexity dependence on the variance $\Delta$ becomes
non-monotonic: at values of the non-extensivity parameter limited to the top
magnitude $Q _ +\simeq 1.618$, the maximum complexity increases with $\Delta$
growth, whereas at $Q> Q _ + $ it decreases. With accounting for the
self-similarity condition (\ref {10}), this means that behavior inherent in
simple statistical systems is realized at the branching exponent of
hierarchical tree exceeding gold mean $a_+\equiv(\sqrt {5} +1)/2\simeq 1.618$;
on the other hand, the complexity decay with the variance growth, what is
characteristic for complex systems, is shown to appear at values of the
branching exponent within the domain $1<a <1.618$.

\section{Discrete hierarchical levels}\label{sec:4}

\begin{figure}[!htb]
\centering
\includegraphics[width=80mm]{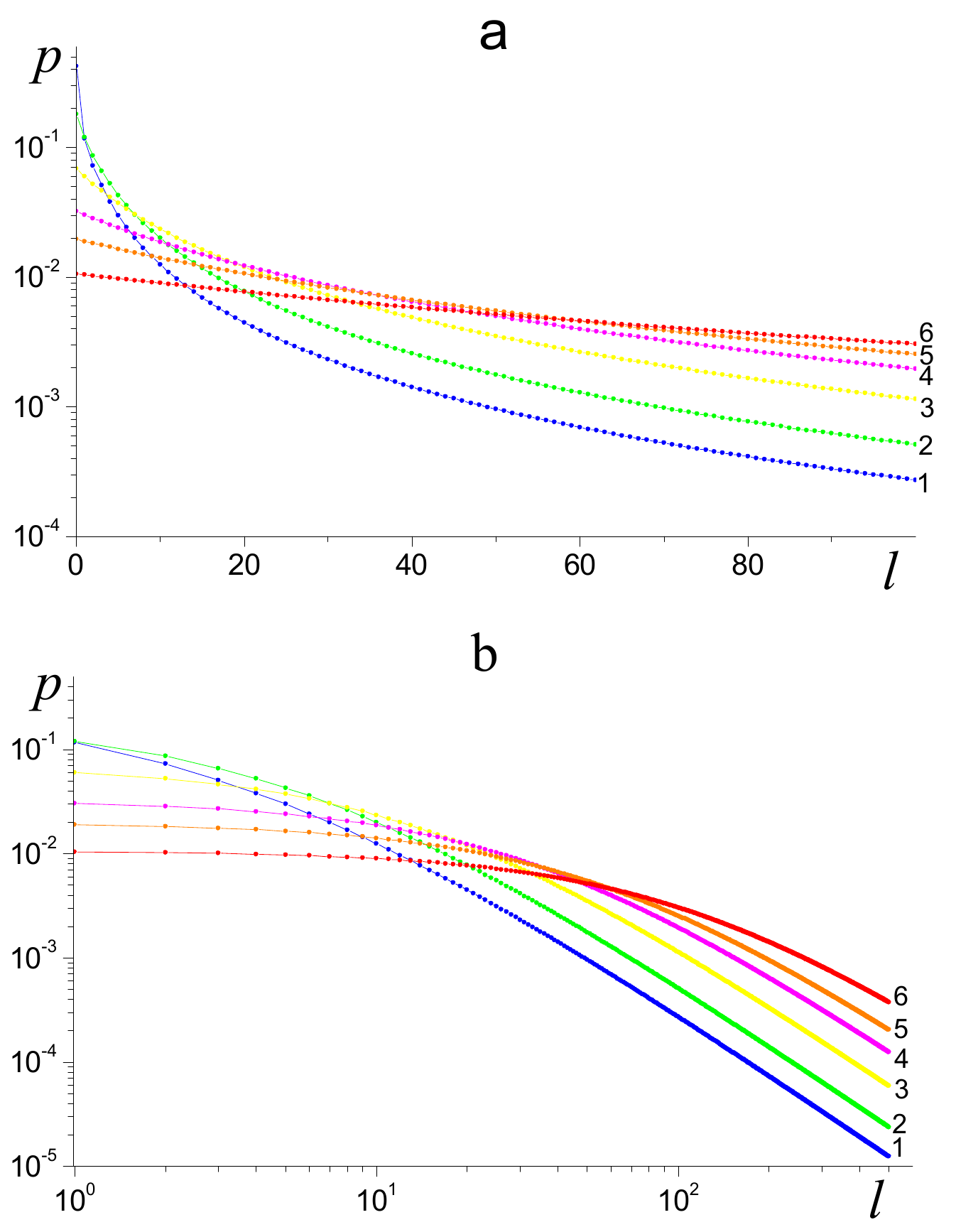}
\caption{ Probability distributions over hierarchical levels within
representations of both half-logarithmic (a) and log-log (b) axes at $Q=1.5$
(curves 1 -- 6 relate to values $\Delta=0.9,1.25,2.0,3.0,4.0,6.0$,
respectively).}\label{fig2}
\end{figure}

For hierarchical systems possessing a finite number of levels $n> 1$, the above
results of the continuum approach lose accuracy, and we need to take into
account the discretization of hierarchical statistical ensemble. Probability
distribution of such ensemble is defined by the set of equations (\ref{11}),
whose number $n$ is less by one than number of probabilities $p_l$,
$l=0,1,\dots,n$. The ensemble description is reached by adding to the system
(\ref {11}) the normalization condition
\begin{equation}
\sum\limits _ {l=0} ^ {n} p_l=1.
 \label {23}
\end{equation}
Numerical solution of Eqs. (\ref {11}), (\ref {23}) arrives at the probability
distributions over hierarchical levels with different variances, as shown in
Fig.\ref{fig2} (for convenience we
use both half-logarithmic axes, where exponential dependencies are
straightened, and logarithmic axes, representing power laws with straight
lines). As Fig.\ref{fig2}a shows, distributions of statistical states over
hierarchical levels are far from exponential dependencies at all values of the
variance $\Delta$. The pronounced curve straightening in Fig.\ref{fig2}b
indicates the transition into power-law regime in the limit $l\to\infty$. The
latter is confirmed by comparing the distribution
\begin{figure}[!htb]
\centering
\includegraphics[width=80mm]{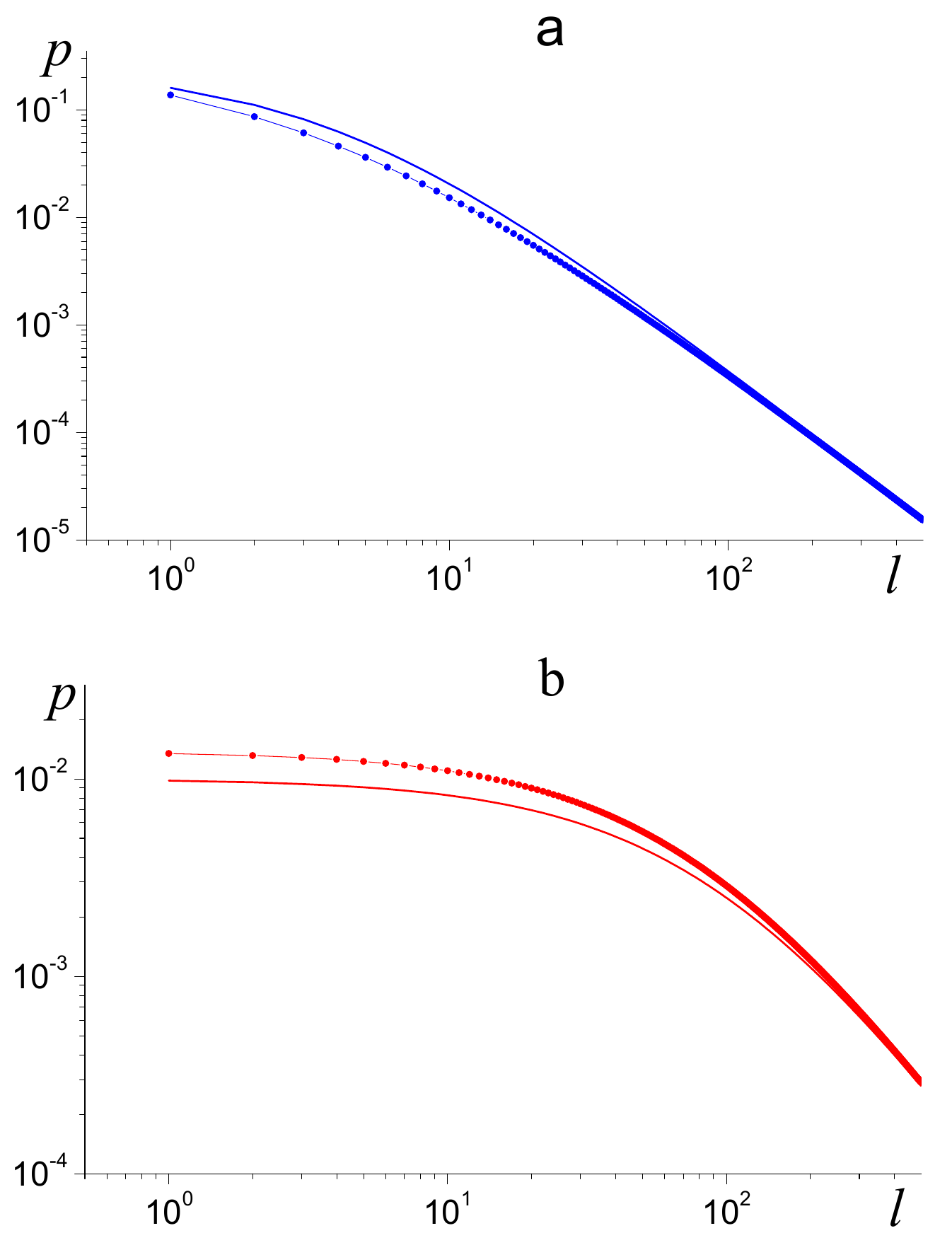}
\caption{Comparison of the distribution function (\ref {2}) found within
continuum approach (solid lines) with solution of the difference equations
(\ref{11}) and the normalization condition (\ref {23}) (points) at $Q=1.5$,
$\Delta=1$ (a) and $Q=1.5$, $\Delta=5$ (b).}\label{fig3}
\end{figure}
function (\ref{2}) derived in the continuum limit with the solution of the
system of difference equation (\ref {11}) subject to the normalization
condition (\ref {23}). As illustrated in Fig.\ref{fig3}, both approaches
produce virtually the same results for the deep levels $l\gg 1$, however, the
discrepancy at lower level numbers becomes bigger as the variance $\Delta$
increases.

\begin{figure}[!htb]
\centering
\includegraphics[width=80mm]{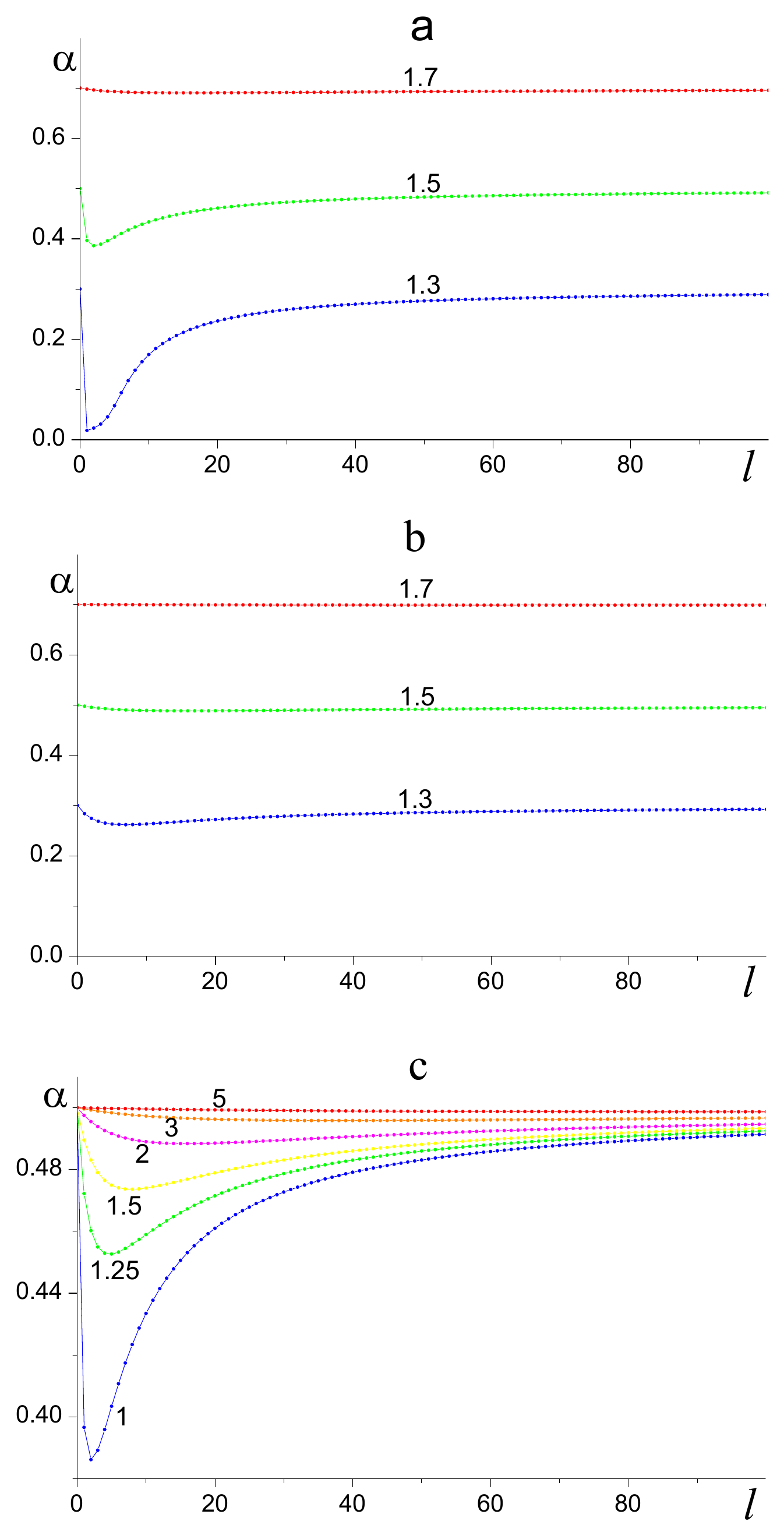}
\caption{Dependencies of the exponent of distribution (\ref {24}) on the level
number at $\Delta=1$ (a), $\Delta=2$ (b) and $Q=1.5$ (c) (numbers near curves
point to values $Q$ in panels (a), (b) and $\Delta$ in (c)).}\label{fig4}
\end{figure}

The most precise representation of accuracy of the continuum approach is
reached if one expresses the distribution over hierarchical levels in the
modified Tsallis form
\begin{equation}
\begin{split}
&p_l=\left [p_0 ^ {-(Q-1)} + \frac {\alpha_l} {\Delta} l\right] ^
{-{1\over\alpha_l}}; \\ &p_0\equiv \left (\frac {2-Q} {\Delta} \right) ^ {\frac
{1} {2-Q}}, \quad 0\leq l\leq n \label {24}
\end{split}
\end{equation}
with the exponent $\alpha_l$ taking the value $\alpha_l\simeq Q-1$ in the limit
$l\to\infty $. As Fig.\ref{fig4} shows, the continuum approach is improved by
both the tendency of the non-extensivity parameter $Q$ to the limit value $Q=2$
and the unlimited growth of the variance $\Delta$. In particular, $Q\to 2$
means that branching exponent (\ref {10}) tends to the limit $a=1$ related to
the degenerate hierarchical tree, where the statistical ensemble is defined by
hyperbolic distribution with the exponent $\alpha_l\simeq Q-1\to 1$.
Respectively, growth of the variance increases the disorder of statistical
ensemble dramatically and the discrete nature of the ensemble becomes
inessential.

\begin{figure}[!htb]
\centering
\includegraphics[width=80mm]{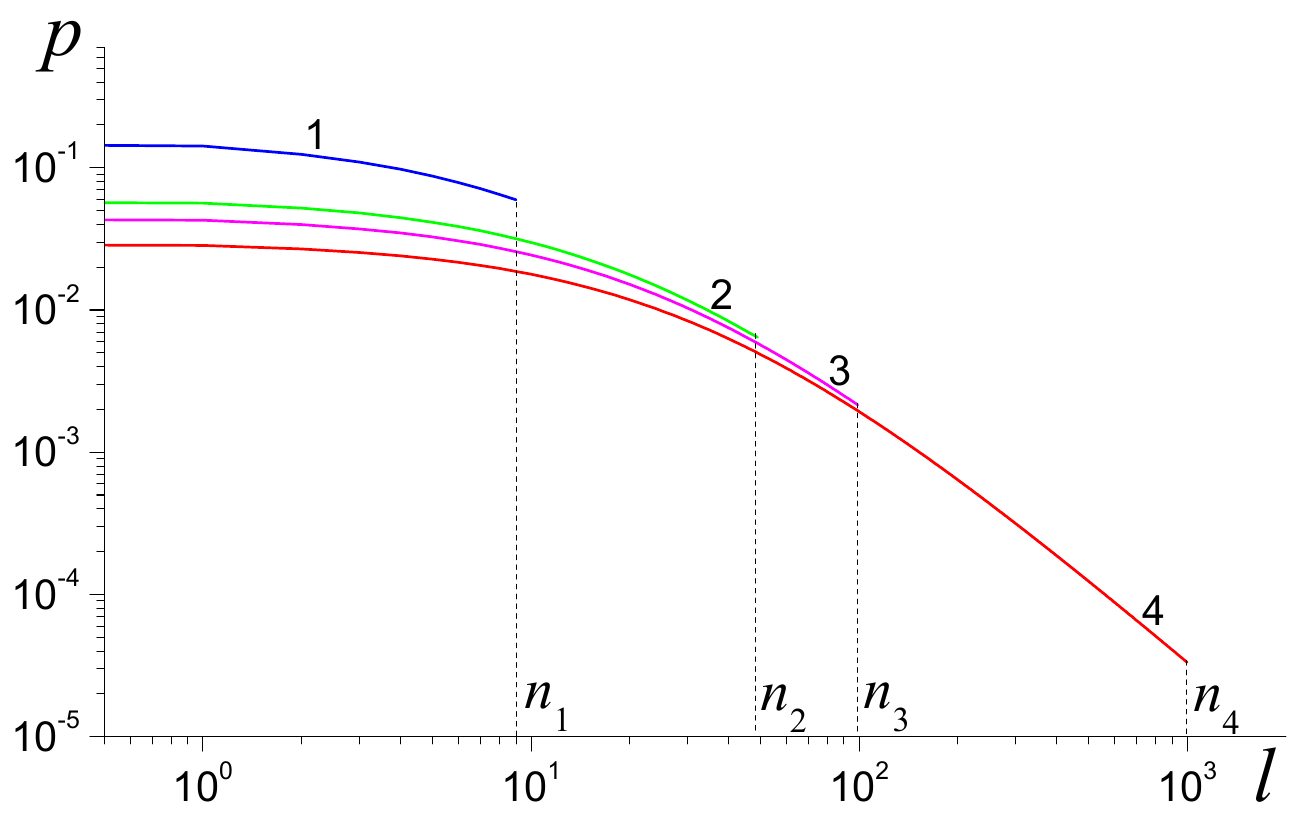}
\caption{Evolution of the probability distribution over hierarchical levels at
$Q=1.5$ and $\Delta=3$ (curves 1 -- 4 relate to the time moments $t/\tau_0=50,
1.25\cdot10^3, 5\cdot10^3, 5\cdot10^5$, respectively).}\label{fig5}
\end{figure}

As in stationary case, evolution of the probability distribution over
hierarchical levels is determined by the system of equations (\ref {11}) along
with the normalization condition (\ref {23}). The number of the deepest level
is defined by the diffusion law $n =\sqrt {2 (t/\tau_0)} $ (for the sake of
simplicity, we accept the condition $\tau_d/\tau\ll 1$ for characteristic time
scales, when the contribution of abnormal drift is negligible). Corresponding
time dependencies of the probability distribution functions are depicted in
Fig.\ref{fig5}.
It is demonstrated how diffusion process over levels of the hierarchical tree
provides relaxation to the stationary distribution shown in Fig.\ref{fig3}.

\begin{figure}[!htb]
\centering
\includegraphics[width=80mm]{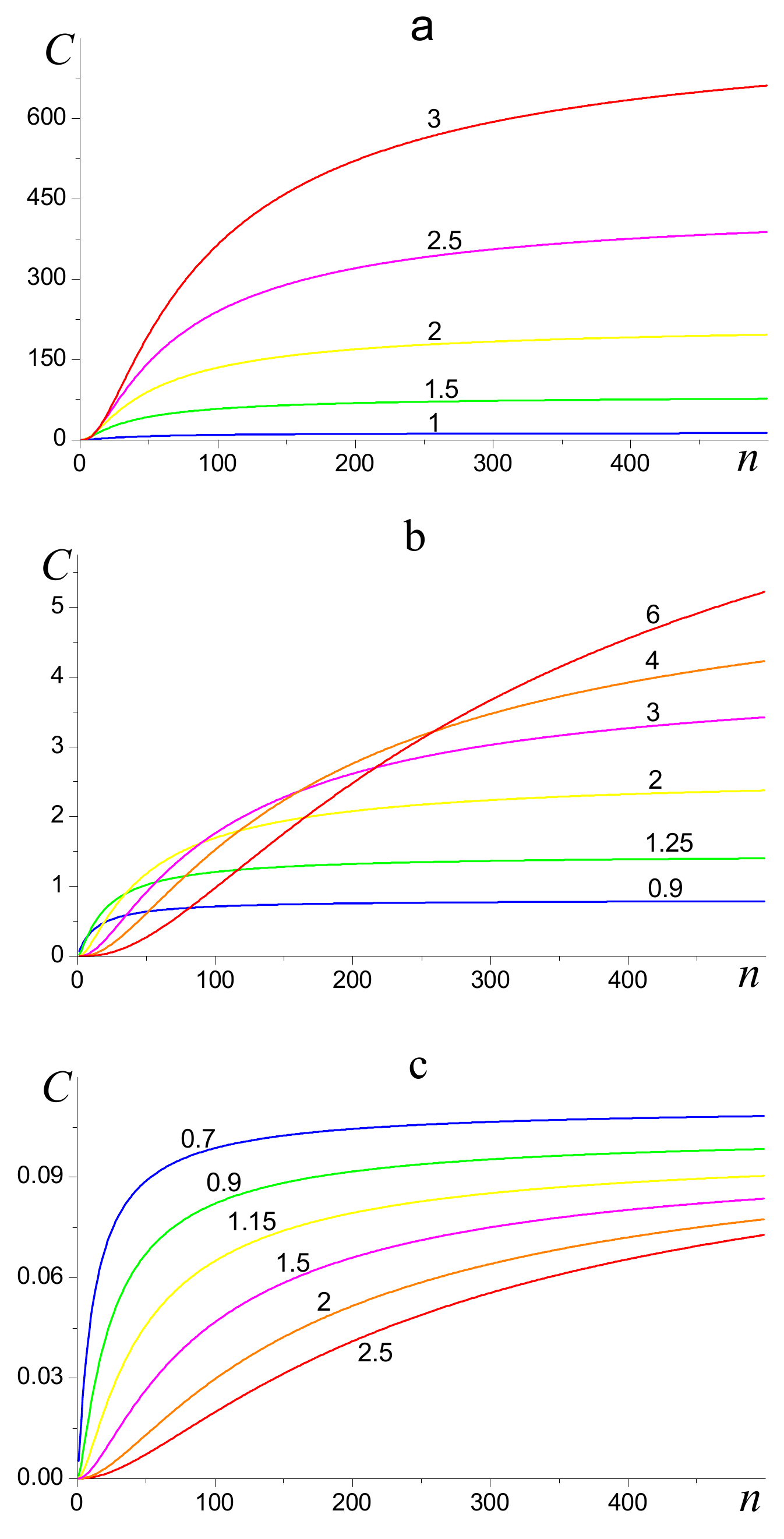}
\caption{Dependence of the complexity on the number of hierarchical levels at
$Q=1.3$ (a), $Q=1.5$ (b) and $Q=1.7$ (c) (numbers near curves point to the
$\Delta$ values).}\label{fig6}
\end{figure}

\begin{figure}[!htb]
\centering
\includegraphics[width=80mm]{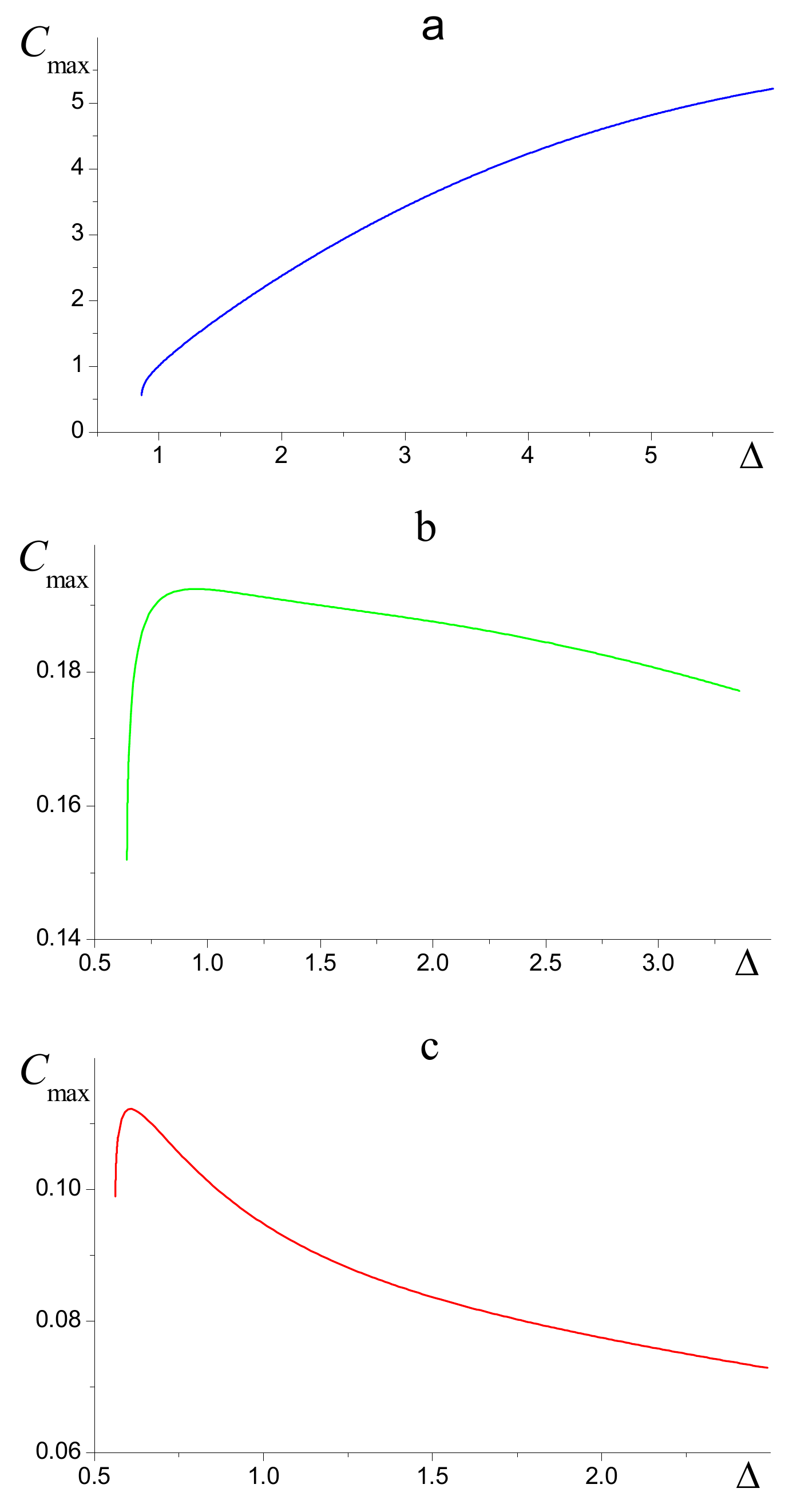}
\caption{Dependence of complexity on variance for levels $n=500$ at $Q=1.50$
(a), $Q=1.65$ (b) and $Q=1.70$ (c).}\label{fig7}
\end{figure}

Let us proceed now with definition of the complexity. Taking successfully
values $l=0,1, \dots, n $ for the number of levels in Eq.(\ref {7}), with
accounting for Eqs. (\ref {8}), (\ref {10}), one obtains
\begin{equation}
C(n) = \frac {1} {Q-1} \sum _ {l=1} ^ {n} l ^ {1 / (Q-1)} p _ {l} \left (p _
{l-1} ^ {Q-1}-p _ {l} ^ {Q-1} \right).
 \label {25}
\end{equation}
As shown in Fig.\ref{fig6}, with strengthening hierarchical coupling the
complexity grows monotonically to the maximum value which decays to zero as the
non-extensivity parameter $Q $ tends to the upper border $Q=2$. What about the
complexity dependence on the variance $\Delta$, it is much more complicated: at
small values of $Q$ the $\Delta$-growth promotes the complexity-increase,
whereas at large $Q$ the complexity decreases with the variance. Fig.\ref{fig7}
demonstrates this peculiarity
representing the complexity as a function of variance $\Delta$ for a large
number of levels $n=500$ and different values of $Q$. It is seen that
complexity increases monotonically with $\Delta$ at small $Q$, decays at large
$Q$ and varies non-monotonically at intermediate $Q$.

According to consideration of the previous Section, such complexity behavior is
also captured by the relation (\ref {21}). The latter, in particular, implies
that the maximum complexity as a function of variance exhibits the behavior
inherent in simple statistical systems at values of non-extensivity parameter
limited from above by the gold mean $Q _ +\simeq 1.618$. On the other hand, at
$Q>Q_+$ the variance-growth reduces the complexity of hierarchical ensemble.
Qualitatively, such complexity dependence can be perceived while considering a
trivial hierarchical tree (see Fig.\ref{fig1}b) in line with the following
reasonings. As the top node splits into $m>1$ parallel branches, each of which
comprises of $n>1$ sub-nodes, the total number of nodes equals $N=1+mn $; at
that, the top node relates to the probability $p_0=m/N $, whereas each of the
remaining sub-nodes is associated with the probability $p=1/N $. Then, the
complexity definition (\ref {25}) gives the value $C_1 = (Q-1) ^ {-1} p (p_0 ^
{Q-1}-p ^ {Q-1}) $ for each of the tree branches and the total complexity
$C=mC_1$ takes the value
\begin{equation}
C = (Q-1) ^ {-1} \frac {1-m ^ {-(Q-1)}} {(n+m ^ {-1}) ^Q} \sim n ^ {-Q}.
 \label {25a}
\end{equation}
As one would expect, the latter grows with increase of the branching $m $ and
reduction of the length of branches $n $. Since for a stochastic tree the
$n$-growth is attributed to the increase in disorder of states over levels, the
estimation (\ref {25a}) provides an explanation for an abnormal decay of the
complexity with growth of the variance of hierarchical ensembles -- it is
caused by the increase in number of subensembles which are connected with each
other without hierarchical constraining.

Characteristic peculiarity of the curves depicted in Fig. \ref{fig7} is their
break at small variance, where the complexity (\ref {25}) becomes ill-defined.
Such behavior is caused by the functional form of the distribution (\ref {2}),
implying that the decrease in variance $\Delta$ leads to unlimited growth of
the probability to occupy the top level $p_0\propto \Delta ^ {-1 / (2-Q)} $.
Therefore, at small $\Delta$ the maximum probability $p_0=1$ is reached, so the
whole hierarchical system accumulates on the top level and the definition of
complexity loses sense.

\begin{figure}[!htb]
\centering
\includegraphics[width=80mm]{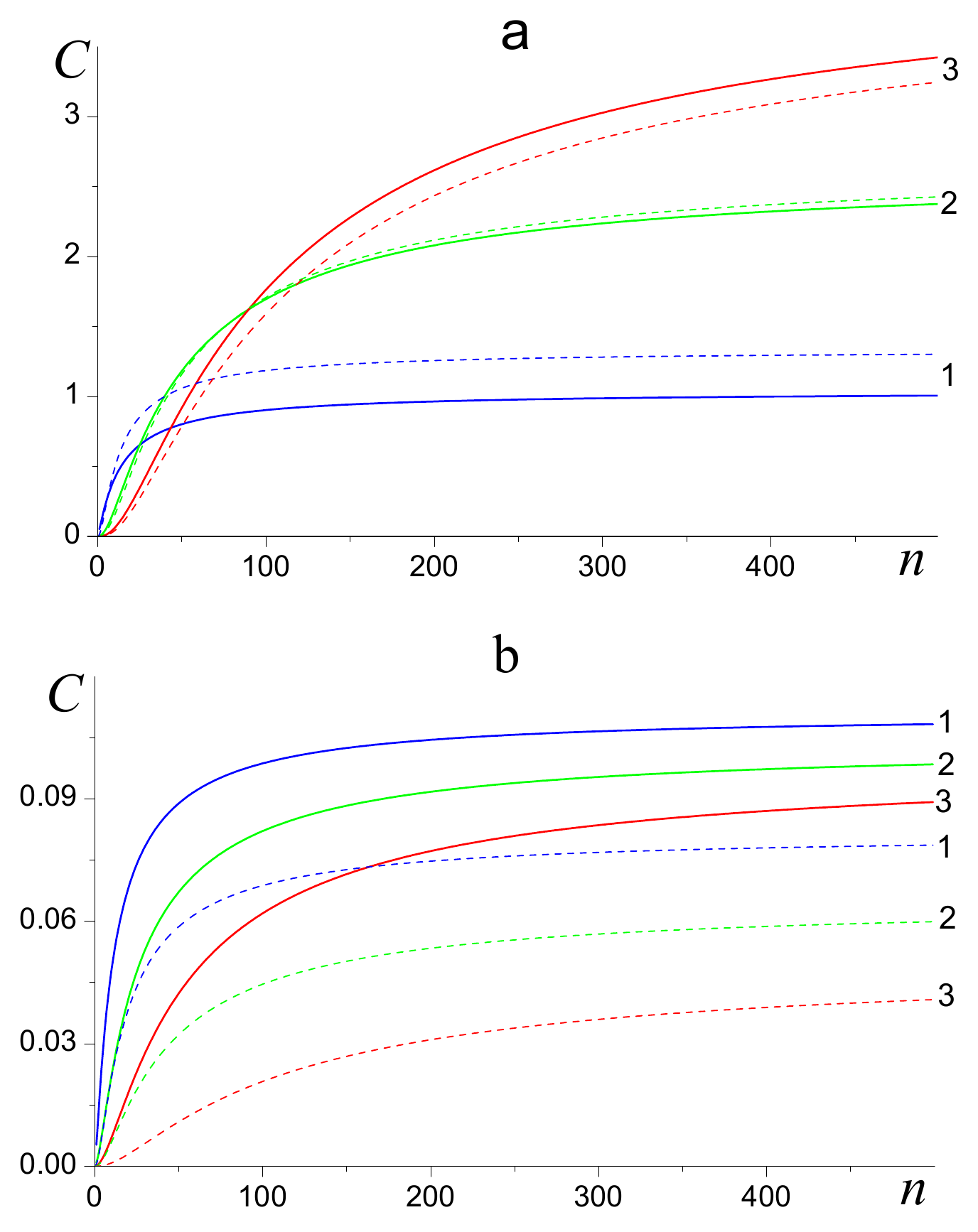}
\caption{Comparison of complexities (\ref {25}) determined within both discrete
(solid lines) and continuum (dashed lines) approaches: a) $Q=1.5$ (curves 1 --
3 relate to $\Delta=1,2,3$, respectively); b) $Q=1.7$ (curves 1 -- 3 relate to
$\Delta=0.7,0.9,1.2$, respectively).}\label{fig8}
\end{figure}

Comparison between the complexities (\ref {25}) and (\ref {20}) defined within
discrete and continuum approaches, respectively, is shown in Fig.\ref{fig8}.
It is seen that both approaches result in the identical complexity dependence
on the number of hierarchical levels $n$. However, with growth of the
non-extensivity parameter $Q$ the continuum approach loses the accuracy. Such
behavior is rationalized by the reduction of the tree branching with
$Q$-growth, promoting enhanced contribution of top levels where the continuum
approach is inapplicable.

\begin{figure}[!htb]
\centering
\includegraphics[width=80mm]{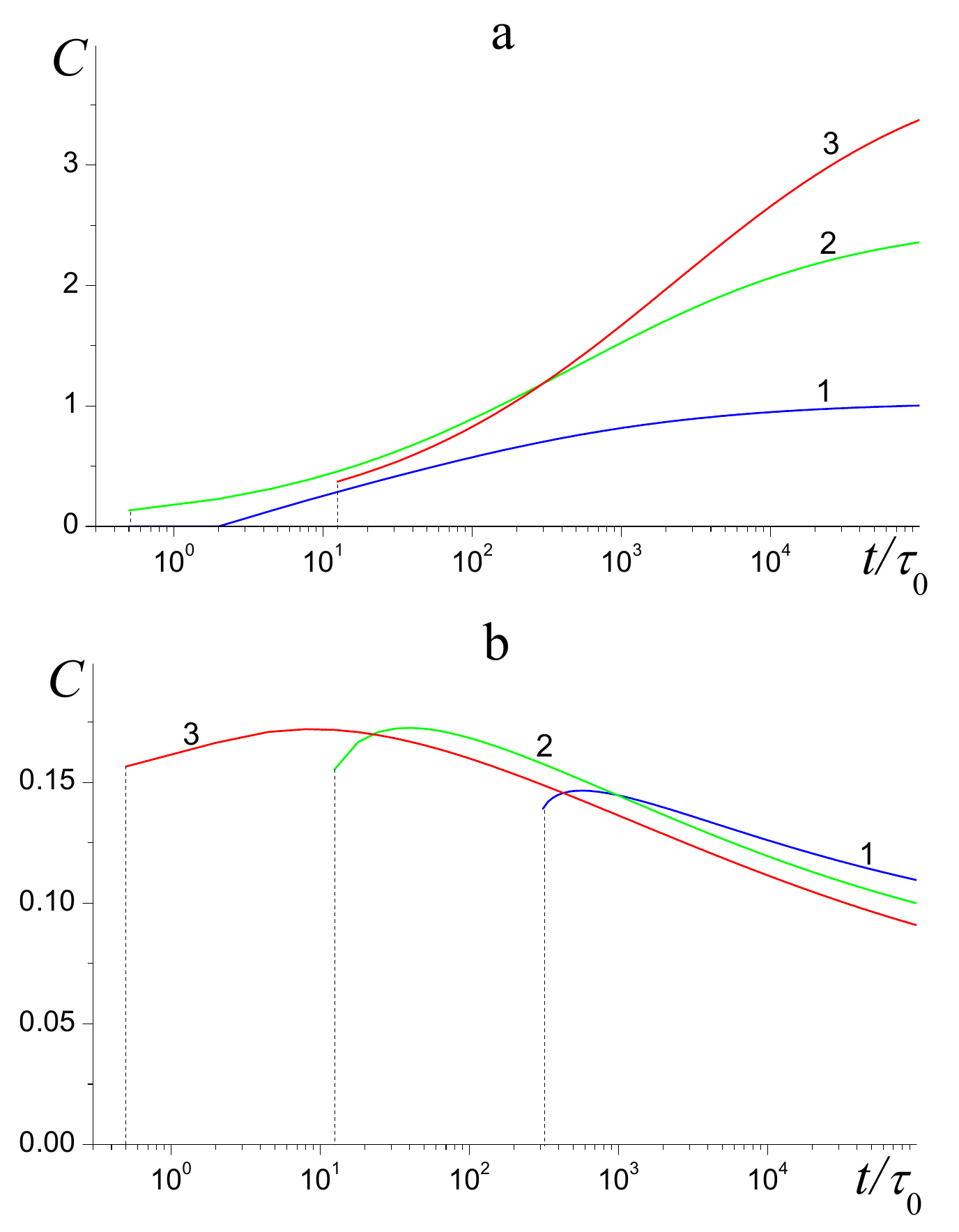}
\caption{Time dependencies of the complexity at $Q=1.5$ (a) and $Q=1.7$ (b);
curves 1 -- 3 relate to values $\Delta=1, 2, 3$ (a) and $ \Delta=0.7, 0.9, 1.2$
(b), respectively.}\label{fig9}
\end{figure}

Time dependence of the complexity $C(t)$ is defined by evolution of the
probability distribution $p_l(t)$ over hierarchical levels. Fig.\ref{fig5}
shows that in the course of time the level variance is strengthened, so,
accordingly, the dependence $C(t)$ should have the same form as the complexity
variation with the $\Delta$ variance growth for stationary case (see
Fig.\ref{fig7}). Indeed, as indicated in Fig.\ref{fig9}, complexity increases
with time the faster, the greater the distribution scattering, provided the
non-extensivity parameter does not exceed the gold mean $Q _ +\simeq 1.618$
\cite {9}.
However, at $Q> Q _ + $ the system evolution becomes abnormal: a short-term
increase of the complexity from finite values is followed by its long-continued
decay to the values decreasing with the $\Delta$ variance growth.

\section{Discussion}\label{sec:5}

Above consideration is based on the definition of microcanonical ensemble,
where the probabilities $p _ {i_1\dots i_l} $ to occupy statistical states
$i_1\dots i_l $ on $l$th hierarchical level take equal values $p_l\sim p_0/M_l$
defined by the condition (\ref {9}). According to Eq.(\ref {8}), these
probabilities decay with $l$-growth due to increase in the number of states
$M_l$. As a result, the recurrence relation (\ref {6}) is reduced to simple
equality (\ref {7}), whose iteration gives the expression (\ref {25}) for the
complexity of the self-similar hierarchical ensemble.

In general case of arbitrary distribution over nodes of the hierarchical tree,
not possessing the property of self-similarity, the difference equation (\ref
{6}) leads to the complicated expression
\begin{equation}
\begin{split}
 &C (n) = \frac {1} {Q-1} \sum _ {l=1} ^ {n} C_l, \\ C_l\equiv & \sum _
{i_1=1} ^ {m} \sum _ {i_2=1} ^ {m _ {i_1}} \dots\sum _ {i_l=1} ^ {m _
{i_1i_2\dots i _ {l-1}}} p _ {i_1\dots i _ {l-1} i_l}\times \\ &\left (p _
{i_1\dots i _ {l-1}} ^ {Q-1}-p _ {i_1\dots i _ {l-1} i_l} ^ {Q-1} \right),
\label {26}
\end{split}
\end{equation}
where the probability $p _ {i_1\dots i _ {l-1}} $ is reduced to the value $p_0$
at $l=1$. In contrast to a simple two-level tree depicted in Fig.\ref{fig1}à,
it is meant here that on the level $l$ of the hierarchical tree comprised of
$n>1$ total number of levels the set of states $i_1\dots i _ {l-1} 1$,
$i_1\dots i _ {l-1} 2$, $ \dots $, $i_1\dots i _ {l-1} m _ {i_1i_2\dots i _
{l-1}} $ forms a cluster related to the node $i_1\dots i _ {l-1} $ on the upper
level $l-1$ (see Fig.\ref{fig1}c). Therefore, to calculate the complexity (\ref
{26}), one needs first to sum over the nodes $i_1\dots i _ {n-1} i _ {n} $ of
the bottom level $n $, which belong to the cluster related to the node
$i_1\dots i _ {n-1} $ of the upper level $n-1$. Then, the summation over nodes
of the level $n$ belonging to all remaining clusters is carried out and the
same procedure is repeated for each subsequent hierarchical level $l<n$
\footnote {At complexity definition of lightly branching trees, it is more
convenient to carry out summation not over clusters, but over branches of the
hierarchical tree.}.

Expression (\ref {26}) is a basis for the numerical definition of the
complexity of arbitrary hierarchical ensemble (for example, in case of the
complex defect structure of solids subject to intensive external influence type
of strong plastic deformation or rigid radiation treatment). Unlike the
amorphous systems, the number of structure levels of a real crystal is rather
not large. Particularly, among different spatial scales in the latter type of
systems, it is accepted to distinguish between several basic levels of
consideration \cite{15}. The finest level, the microscopic one, is related to
homogeneous distribution of point defects, dislocation and disclination. Next,
the coarser mesoscopic level corresponds to homogeneously distributed structure
cells and fragments. Finally, the macroscopic objects such as homogeneously
distributed grains and texture components are considered on a macroscopic
level. To define the complexity of a real structure, one needs first to
distribute the whole ensemble of defects over hierarchical levels $l=0,1,
\dots, n $, then to calculate on each of them a number of defects $N _
{i_1\dots i _ {l-1}} $ belonging to the cluster $i_1\dots i _ {l-1} 1$,
$i_1\dots i _ {l-1} 2$, $ \dots $, $i_1\dots i _ {l-1} m _ {i_1i_2\dots i _
{l-1}} $, and, finally, to attribute the probability
\begin{equation}
p _ {i_1\dots i _ {l-1}} = \frac {N _ {i_1\dots i _ {l-1}}} {N}
 \label {27}
\end{equation}
to the node $i_1\dots i _ {l-1} $ of the upper level $l-1$. At that, the total
number of defects residing on all levels is defined as
\begin{equation}
N=\sum _ {i_1=1} ^ {m} \sum _ {i_2=1} ^ {m _ {i_1}} \dots\sum _ {i _ {n} =1} ^
{m _ {i_1\dots i _ {n-1}}} N _ {i_1\dots i _ {n}}, \quad n> 1 \label {28}
\end{equation}
where $m _ {i_1\dots i _ {l-1}} $ is the number of possible states $i_1\dots
i_l $ within the cluster related to the node $i_1\dots i _ {l-1} $ (at $n=1$
one has $m _ {i_1\dots i _ {n-1}} =m $) \footnote{In general case, distribution
of states $m _ {i_1\dots i _ {l}} $ over all clusters defines their number on
given level $l $ according to the equality $$M_l=\sum _ {i_1=1} ^ {m} \sum _
{i_2=1} ^ {m _ {i_1}} \dots\sum _ {i _ {l} =1} ^ {m _ {i_1\dots i _ {l-1}}} m _
{i_1\dots i _ {l}}.$$ For regular tree where each of nodes branches with
constant exponent $m>1$, one has from here $M_l=m^l\equiv\exp\left (\ln m\cdot
l\right) $. Passage to irregular self-similar tree transforms this expression
to binomial dependence $M_l =\left [1 + (\ln m/a) \cdot l\right] ^a $, which
reproduces above exponential in the limit $a\to\infty $ and is reduced to the
power-law (\ref {8}) at the exponent $a =\ln m $.}. Substitution of the
obtained set of probabilities $ \left \{p _ {i_1\dots i_l} \right \} $, $l=0,1,
\dots, n $ into Eq.(\ref {26}) gives the complexity of hierarchically
constrained defect structure of a solid. Obviously, this complexity determines
such phenomenological quantities as strength and plasticity of solid.
Remarkably, such definition of the structural complexity cannot be reached
solely by means of experimental methods (for example, electron microscopy), but
requires a subsequent computer processing according to the above algorithm.

Realization of the described program, however, represents rather a challenge.
Therefore, in the preceding sections we limited ourselves to considering a
self-similar hierarchical tree characterized by the level number $n$ and a
typical cluster with the branching exponent $a$. Studying the various
hierarchical trees has revealed three their types \cite {12}: (1) the
degenerate tree possessing only one branching node per level, so the total
number of nodes increases linearly with $n$; (2) the regular tree, where on
each level all nodes branch equally, and the total node number increases
exponentially with $n$; and (3) the self-similar irregular tree characterized
by the power law (\ref{8}) with exponent $a>1$ for the $l$-dependence of the
number of nodes. In the first two cases the probabilities of state occupation
on various hierarchical levels vary logarithmically slow and exponentially
fast, respectively, while in the last case it obeys the power-law dependence
(\ref {2}) inherent in the self-similar systems. According to Eq.(\ref {10}),
the non-extensivity parameter $Q = (a+1)/a $ is determined by the branching
exponent $a$: regular $(a=\infty)$ and degenerate $(a=1)$ trees are
characterized by limiting values $Q=1$ and $Q=2$, respectively, whereas the
power-law distribution (\ref {2}) with exponent $1 <Q <2$ takes place at
$\infty>a>1$.

As pointed in Introduction, the complexity of different hierarchical trees has
been first considered in works \cite {10a}, \cite {9a}. Within our notation,
these trees have been characterized with the {\it silhouette}
\begin{equation}
s_l:=\ln\frac{M_{l}}{M_{l-1}}
 \label{30}
\end{equation}
that determines the logarithmic growth rate of the node number $M_l$ with
increase of the hierarchical level number $l$. A peculiarity of the approach
used in Ref.\cite {10a} is that it is based on the consideration of the regular
tree, whose nodes are multifurcated with the constant branching index $b>1$. In
this case, one has the dependence $M_l=b^l$ and the definition (\ref {30})
gives a simple relation $s=\ln b$. However, for the main object of our
consideration, a self-similar tree, Eq. (\ref {8}) for the number node
distribution has to be used, and the resulting silhouette
$s_l=\ln\left(1+1/l\right)^a\simeq a/l$ depends significantly on the level
number even in the continuum limit $l\gg 1$. It is clear that such
$l$-dependence is caused by the definition (\ref {30}) introduced for trees
related to the regular ones, whereas self-similar irregular trees represent
their antipode. Obviously, the silhouette of a self-similar tree may be defined
with the equalities
\begin{equation}
s:=\frac{M_{\lambda l}-M_l}{(\lambda-1)M_l}=[a]_\lambda,\quad
[a]_\lambda\equiv\frac{\lambda^a-1}{\lambda-1}.
 \label{31}
\end{equation}
This definition is based on use of the Jackson derivative which determines the
variation rate of the function (\ref {10}) with respect to the dilatation
$\lambda\geq 1$ and reduces to the usual derivative in the limit $\lambda\to 1$
\cite{Jackson}. According to Eq.(\ref {30}), the silhouette of self-similar
tree is the $\lambda$-basic number $[a]_\lambda$, whose value equals the
exponent $a$ at the limit dilatation value $\lambda=1$ and grows as
$\lambda^{a-1}$ at $\lambda\gg 1$. Obviously, the present study of self-similar
hierarchical ensembles relates to the first case{\bf, $\lambda=1$}.

In summary, our study indicates that formation of hierarchical coupling
promotes a fast complexity growth to a certain maximum value. With
strengthening of the branching parameter $a$ of hierarchical tree the maximum
complexity grows monotonically from zero at $a=1$ to infinity at $a\to\infty$.
Investigating the variance-dependence of the complexity has revealed a behavior
inherent in simple systems at the values of branching exponent exceeding the
gold mean $a _ + = 1.618$. On the other hand, the complexity-decrease with the
variance-growth, being a characteristic of complex systems, has been observed
for the range $1<a <1.618$ of the branching exponent.

\begin{widetext}
\section*{Appendix}\label{sec:6}
 \def\theequation{{A}.\arabic{equation}}
 \setcounter{equation}{0}

As known, the non-extensive statistical mechanics is based on the following
definitions for logarithmic and exponential functions \cite{11},
\begin{equation}
\ln_q(x):= \frac{x^{1-q}-1}{1-q},\ \exp_q(x):= \left[1+(1-q)x\right]_+^{1\over
1-q};\quad [y]_+:= \max(0,y),\ q\leq 1,
 \label{A1}
\end{equation}
which are reduced to the usual functions in the limit $q\to 1$. Introducing the
$q$-deformed multiplication and division operations for the positive values $x,
y$ as follows,
\begin{equation}
x\otimes_q y=\left[x^{1-q}+y^{1-q}-1\right]_+^{1\over 1-q},\ x\oslash_q
y=\left[x^{1-q}-y^{1-q}+1\right]_+^{1\over 1-q};\quad x,y>0,
 \label{A2}
\end{equation}
it is easy to verify that they satisfy the usual properties
\begin{equation}
\begin{split}
&\ln_q(x\otimes_q x)=\ln_q x+\ln_q y,\quad \ln_q(x\oslash_q x)=\ln_q x-\ln_q
y;\\ &\exp_q(x)\otimes_q\exp_q(y)=\exp_q(x+y),\quad
\exp_q(x)\oslash_q\exp_q(y)=\exp_q(x-y)
\end{split}
\end{equation}
of the logarithmic and exponential functions.

Within combinatorial approach \cite{13}, the $q$-deformed statistics is reduced
to consideration of the generalized factorial $N!_q:=1\otimes_q\dots\otimes_q
N$ and the corresponding logarithm,
\begin{equation}
\ln_q(N!_q)=\frac{\sum_{i=1}^{N}i^{1-q}-N}{1-q}.
 \label{A3}
\end{equation}
In the thermodynamic limit $N\to\infty$, a sum in the above equation is
replaced by an integral, and one gets
\begin{eqnarray}
\ln_q(N!_q)=\left\{
\begin{array}{ll}\frac{N}{2-q}\ln_q N-\frac{N}{2-q}+{\it O}(\ln_q N),\qquad q\ne 2,\\
N-\ln N+{\it O}(1),\quad\quad\quad\quad\quad\quad\quad q=2.
\end{array} \right.
\label{A4}
\end{eqnarray}
Defining $q$-deformed multinomial coefficient as
\begin{equation}
{N\choose N_1\dots N_k}_q
:=(N!_q)\oslash_q\left[(N_1!_q)\otimes_q\dots\otimes_q(N_k!_q)\right]
 \label{A5}
\end{equation}
with a set of integers $N_i$ subject to the condition $N=\sum_{i=1}^{n}N_i$, we
find
\begin{equation}
{N\choose N_1\dots N_k}_q
=\left[\sum\limits_{i=1}^{N}i^{1-q}-\sum\limits_{i_1=1}^{N_1}i_1^{1-q}-\dots
-\sum\limits_{i_k=1}^{N_k}i_k^{1-q}+1\right]_+^{1/(1-q)}.
 \label{A6}
\end{equation}
Similarly to Eq.(\ref{A4}), the logarithm
\begin{eqnarray}
\ln_q{N\choose N_1\dots N_k}_q\simeq\left\{
\begin{array}{ll}\frac{N^{2-q}}{2-q}C_{2-q}\left(\frac{N_1}{N},\dots,\frac{N_k}{N}\right),\qquad q>0,q\ne 2,\\
-C_1(N)+\sum\limits_{i=1}^{k}C_{1}(N_i),\quad\quad\quad\quad\quad\quad q=2
\end{array} \right.
\label{A7}
\end{eqnarray}
is calculated to construct a Tsallis entropy
\begin{equation}
C_Q(p_1,\dots,p_N):=-\sum\limits_{i=1}^{N}p_i\ln_{2-Q}(p_i)=\frac{\sum_{i=1}^{N}p_i^{Q-1}-1}{Q-1}
 \label{A8}
\end{equation}
related to the physical parameter $Q\equiv 2-q\geq 1$.

The above formalism can be easily generalized to study hierarchical systems
\cite{16}. To this end, let us consider a structure of the hierarchical
ensemble comprised of $N$ states. These states are distributed over
subensembles $i=1,\dots,m$, each containing $N_i$ states. In turn, every of
these subensemble further splits in $m_i$ smaller subensembles $ij$, each
possessing $N_{ij}$ states. As the ensemble states are distributed, the
relations $\sum_{j=1}^{m_i}N_{ij}=N_i$, $\sum_{i=1}^{m}N_i=N$ hold. Then,
expression (\ref{A5}) for the multinomial coefficients takes a generalized
form,
\begin{equation}
{N\choose N_{11}\dots N_{mm_m}}_q={N\choose N_{1}\dots
N_m}_q\otimes_q{N_1\choose N_{11}\dots
N_{1m_1}}_q\otimes_q\dots\otimes_q{N_m\choose N_{m1}\dots N_{mm_m}}_q,
 \label{A9}
\end{equation}
whose $q$-logarithm is
\begin{equation}
\ln_q{N\choose N_{11}\dots N_{mm_m}}_q=\ln_q{N\choose N_{1}\dots
N_m}_q+\sum\limits_{i=1}^{m}\ln_q{N_i\choose N_{i1}\dots N_{im_i}}_q.
 \label{A10}
\end{equation}
Applying the estimation (\ref{A7}) to the last formula arrives at the
connection (\ref{3}) between the complexities of the nearest hierarchical
levels.

\section{Acknowledgements}\label{sec:level7}

We are grateful to Dr. B.A. Huberman for bringing to our attention Refs.
\cite{10a}, \cite{9a} related to the addressed issues. We also thank Dr. A.P.
Savelyev for helpful discussions and careful manuscript reading.
\end{widetext}

\end{document}